\documentclass[biblatex,runningheads,english]{lni}
\usepackage[subtle]{savetrees}
\usepackage{float}  % \begin{figure}[H]
\usepackage{booktabs}  % nicier tables
\usepackage{pifont}
\usepackage{tikz}
\usetikzlibrary{shapes}

\definecolor{Red}{HTML}{EE2967}
\definecolor{Green}{HTML}{3FBC9D}
\definecolor{Orange}{HTML}{FDBC42}

\begin{document}
\title[Reproducible Builds and Insights from an Independent Verifier for Arch Linux]{{Reproducible Builds and Insights\\from an Independent Verifier for Arch Linux}}

% \author[Joshua Drexel \and Esther Hänggi \and Iyán Méndez Veiga]
% {Joshua Drexel\footnote{School of Computer Science and Information Technology, Hochschule Luzern (HSLU),\\ 6343 Rotkreuz, Switzerland.} \and
% Esther Hänggi\footnotemark[1]{} \ and
% Iyán Méndez Veiga\footnotemark[1]{}$^{,}$\footnote{Institute for Theoretical Physics, ETH Zurich, 8093 Zurich, Switzerland.\\Email: \email{iyan.mendezveiga@hslu.ch}}}

\author[1]{Joshua Drexel}{}{}
\author[1]{Esther Hänggi}{}{}
\author[1,2]{Iyán Méndez Veiga}{}{}
\affil[1]{School of Computer Science and Information Technology\\Lucerne University of Applied Sciences and Arts\\6343 Rotkreuz\\Switzerland}
\affil[2]{Institute for Theoretical Physics\\ETH Zurich\\8093 Zurich, Switzerland}

\startpage{1}
\editor{S. Wendzel et al.}
\booktitle{SICHERHEIT 2024}
\yearofpublication{2024}
\lnidoi{10.18420/sicherheit2024_016}
\maketitle

\begin{abstract}
Supply chain attacks have emerged as a prominent cybersecurity threat in recent years. Reproducible and bootstrappable builds have the potential to reduce such attacks significantly. In combination with independent, exhaustive and periodic source code audits, these measures can effectively eradicate compromises in the building process. In this paper we introduce both concepts, we analyze the achievements over the last ten years and explain the remaining challenges. We contribute to the reproducible builds effort by setting up a rebuilder and verifier instance to test the reproducibility of Arch Linux packages. Using the results from this instance, we uncover an unnoticed and security-relevant packaging issue affecting 16 packages related to Certbot, the recommended software to install TLS certificates from Let's Encrypt, making them unreproducible. Additionally, we find the root cause of unreproduciblity in the source code of \texttt{fwupd}, a critical software used to update device firmware on Linux devices, and submit an upstream patch to fix it.
\end{abstract}


% Max 10 pages (w/o appendix & bibliography)
\section{Introduction}
One of the main advantages of Free\footnote{Sometimes called ``libre'' to emphasize that ``free'' should be understood as in freedom, not as \emph{gratis}.}~\cite{gnu-free-software} and Open Source~\cite{open-source-software} Software (FOSS) over proprietary software is the possibility for third parties to independently audit the source code in order to find bugs and backdoors. Programs whose source code is publicly available are not necessarily more secure than closed source alternatives. Indeed, some of the worst cyber incidents in recent years have been linked to bugs affecting open source software \cite{CVE-2014-0160, CVE-2021-44228}. On the other hand, open source projects often attain higher software quality when continuously reviewed by independent developers. This scrutiny not only leads to early detection of bugs and security issues but also fosters a culture of proactive improvement~\cite{raymond1999}. However, even if the source code is secure, this is not sufficient:  Most users and companies using FOSS software do not download the source code and compile it themselves. Instead, they download a binary, a pre-compiled version of the software that can be executed right away. Proving that a certain binary was created from some determined and unmodified source code is a hard task.

A solution to this problem needs to uniquely link source code and binary code, and this is the goal of reproducible builds (R-B)~\cite{reproducible-builds-paper}: Providing an independently-verifiable path from source to binary code. Anyone with access to the source code and detailed building instructions, including dependencies, should be able to transform the thousands or millions of lines of source code into a binary artifact that is bit-by-bit identical every single time the build process is executed. The immediate security benefits of R-B are enormous. For example, independent verifiers can test that distributed binaries have been created from the published and audited source code. Additionally, attacks on the build infrastructure~\cite{linux.org-hacked, freebsd-hacked} can be detected. R-B also bring some benefits beyond security: Identical artifacts lead to higher cache hit ratios, lowering development costs due to more efficient compilations. In the future, R-B could be used in other tasks such as intrusion detection, intelligence gathering using build honeypots, or even become a prerequisite to certify critical software.

R-B still requires a reliable and trusted build toolchain. Even with independent builders and verifiers, if all of them are running the same compromised compiler, the final artifact can include malicious code while being completely reproducible~\cite{repro-builds-msc-thesis}. Obtaining or generating a trusted build toolchain is a hard problem as well. The reason is that it is challenging to start a build process solely from source code. Normally, a specific version of a compiler is built using a pre-compiled binary of another version of the compiler. This leads to a chicken-and-egg problem for trust in the compiler, since we always have to trust at least some initial binary. In the last years, techniques have been developed to increase the trust in pre-compiled compilers such as Diverse Double-Compiling (DDC)~\cite{Wheeler-paper,Wheeler-PhD-thesis}. Bootstrapping, on the other hand, aims to provide ways to generate a working and trusted build toolchain using binary artifacts smaller than a compiler or, ideally, only auditable code. Bootstrappable builds (B-B) is an effort to eliminate the compiler trust loophole in R-B by minimizing, or completely eliminating, the amount of bootstrap binaries required in a build process~\cite{bootstrappable-website}.

In this paper, we will describe what reproducible builds are and the progress which has been made in this topic over the last ten years. We will then report on our contributions to the R-B effort. We have set up a rebuilder and verifier instance to test the reproducibility of Arch Linux official packages. Through this, we have identified and reported a packaging issue affecting 16 packages related to Certbot. Fixing these packages to be reproducible is relevant since this software is the recommended tool to install TLS certificates issued by Let's Encrypt. We have also submitted an upstream patch to fix the root cause of unreproducibility in the source code of \texttt{fwupd}, enabling distributions to provide reproducible packages. The effectiveness of R-B in preventing supply chain attacks heavily relies on the awareness of the developers about this topic. Increasing visibility of R-B is therefore an additional goal of this paper.

\section{Concepts \& Definitions}

\paragraph{Source code} Source code \emph{is the version of software as it is originally written (i.e., typed into a computer) by a human in plain text (i.e., human readable alphanumeric characters)}~\cite{source-code-def}. In other words, the source code of a software is the set of all files that contain human-readable instructions written in a programming language that can be understood by programmers.

\paragraph{Build toolchain} The build toolchain is the set of tools required to transform the source code of a specific programming language into machine code that can be executed by a central processing unit. Depending on the type of the programming language this may include a compiler, an assembler, a linker, an interpreter, etc.

\paragraph{Artifacts} We define artifacts as any output of a build toolchain, with the exception of plaintext logs. This may include machine code executables, libraries, documentation, tarballs, packages, filesystem images, etc.

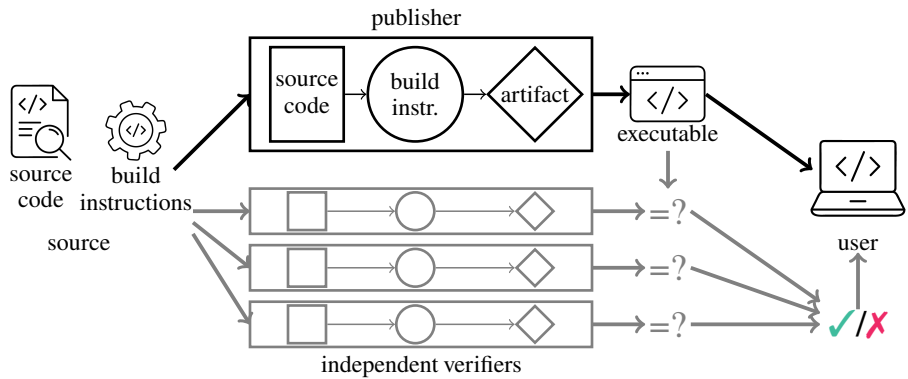
\begin{figure}[H]
\centering
\begin{tikzpicture}	[line width=1pt]
    \path[clip] (-0.7,-2) rectangle (11.2,3);
    \node[align=center, anchor=south] at (-0.25,0.75) {\includegraphics[width=0.8cm]{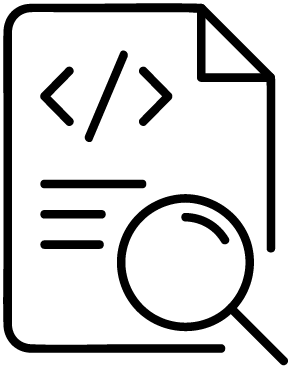}};
    \node[align=center, anchor=center] at (-0.25,0.5) {source\\code};
    \node[align=center, anchor=south] at (1,0.75) {\includegraphics[width=0.8cm]{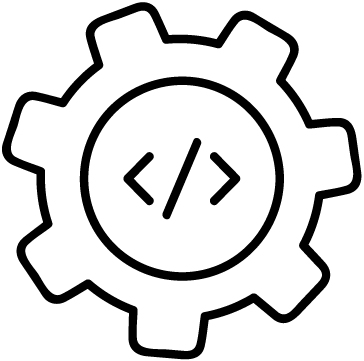}};
    \node[align=center, anchor=center] at (1,0.5) {build\\instructions};
    \node[align=center, anchor=center] at (0.25,-0.25) {source};
    \draw[->, line width=0.5mm, color=black] (1.5,0.75) -- (2.5,1.75);
    \draw [draw=black] (2.5,1) rectangle (7,2.5);
    \node(source)[rectangle, align=center, anchor=center, draw=black, minimum height=1.25cm, inner sep=2pt] at (3.25,1.75) {source\\code};
    \node(build)[circle, align=center, anchor=center, draw=black, minimum height=1.25cm, inner sep=1pt] at (4.675,1.75) {build\\instr.};
    \node(art)[diamond, align=center, anchor=center, draw=black, minimum height=1cm, inner sep=1pt] at (6.25,1.75) {artifact};
    \draw[->, line width=0.2mm, color=black] (source) -- (build);
    \draw[->, line width=0.2mm, color=black] (build) -- (art);
    \node[align=right, anchor=north east] at (5.4,3) {publisher};
    \draw[->, line width=0.5mm, color=gray] (1.75,0.2) -- (2.5,0.2);
    \draw [draw=gray] (2.5,-0.1) rectangle (7,0.5);
    \draw[->, line width=0.5mm, color=gray] (1.75,0.05) -- (2.5,-0.55);
    \draw [draw=gray] (2.5,-0.85) rectangle (7,-0.25);
    \draw[->, line width=0.5mm, color=gray] (1.75,-0.1) -- (2.5,-1.3);
    \draw [draw=gray] (2.5,-1.6) rectangle (7,-1.0);
    \node(source1)[rectangle, align=center, anchor=center, draw=gray, minimum height=0.5cm, minimum width=0.5cm, inner sep=2pt] at (3.25,0.2) {};
    \node(build1)[circle, align=center, anchor=center, draw=gray, minimum height=0.5cm, inner sep=1pt] at (4.675,0.2) {};
    \node(art1)[diamond, align=center, anchor=center, draw=gray, minimum height=0.5cm, minimum width=0.5cm, inner sep=1pt] at (6.25,0.2) {};
    \draw[->, line width=0.2mm, color=gray] (source1) -- (build1);
    \draw[->, line width=0.2mm, color=gray] (build1) -- (art1);
    \node(source2)[rectangle, align=center, anchor=center, draw=gray, minimum height=0.5cm, minimum width=0.5cm, inner sep=2pt] at (3.25,-0.55) {};
    \node(build2)[circle, align=center, anchor=center, draw=gray, minimum height=0.5cm, inner sep=1pt] at (4.675,-0.55) {};
    \node(art2)[diamond, align=center, anchor=center, draw=gray, minimum height=0.5cm, minimum width=0.5cm, inner sep=1pt] at (6.25,-0.55) {};
    \draw[->, line width=0.2mm, color=gray] (source2) -- (build2);
    \draw[->, line width=0.2mm, color=gray] (build2) -- (art2);
    \node(source3)[rectangle, align=center, anchor=center, draw=gray, minimum height=0.5cm, minimum width=0.5cm, inner sep=2pt] at (3.25,-1.3) {};
    \node(build3)[circle, align=center, anchor=center, draw=gray, minimum height=0.5cm, inner sep=1pt] at (4.675,-1.3) {};
    \node(art3)[diamond, align=center, anchor=center, draw=gray, minimum height=0.5cm, minimum width=0.5cm, inner sep=1pt] at (6.25,-1.3) {};
    \draw[->, line width=0.2mm, color=gray] (source3) -- (build3);
    \draw[->, line width=0.2mm, color=gray] (build3) -- (art3);
    \draw[->, line width=0.5mm, color=black] (7,1.75) -- (7.5,1.75);
    \draw[->, line width=0.5mm, color=black] (8.5,1.75) -- (9.9,0.75);
   \node[align=right, anchor=north east] at (6.2,-1.6) {independent verifiers};
       \node[align=center, anchor=south] at (10.5,0) {\includegraphics[width=1.25cm]{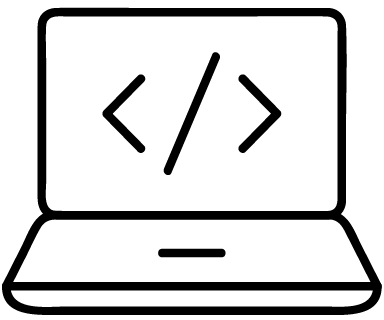}};
    \node[align=center, anchor=center] at (10.5,-0.25) {user};
    \draw[->, line width=0.5mm, color=gray] (8,1.0) -- (8,0.5);
    \draw[->, line width=0.5mm, color=gray] (7,0.2) -- (7.7,0.2);
    \draw[->, line width=0.5mm, color=gray] (7,-0.55) -- (7.7,-0.55);
    \draw[->, line width=0.5mm, color=gray] (7,-1.3) -- (7.7,-1.3);
   \node[align=center, anchor=center,color=gray,inner sep=0pt] at (8,0.2) {\Large{=?}};
   \node[align=center, anchor=center,color=gray,inner sep=0pt] at (8,-0.55) {\Large{=?}};
   \node[align=center, anchor=center,color=gray,inner sep=0pt] at (8,-1.3) {\Large{=?}};
    \draw[->, line width=0.5mm, color=gray] (8.3,0.2) -- (10,-1.0);
    \draw[->, line width=0.5mm, color=gray] (8.3,-0.55) -- (10,-1.15);
    \draw[->, line width=0.5mm, color=gray] (8.3,-1.3) -- (10,-1.3);
    \node[align=center, anchor=center] at (10.5,-1.3) {\Large{\textcolor{Green}{\ding{51}}/\textcolor{Red}{\ding{55}}}};
    \draw[->, line width=0.5mm, color=gray] (10.5,-1.1) -- (10.5,-0.35);
    \node[align=center, anchor=south] at (8,1.25){\includegraphics[width=1cm]{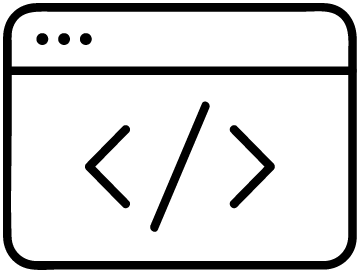}};
    \node[align=center, anchor=south] at (8,1.0){executable};
    \end{tikzpicture}
    \caption{The publisher of an open source software provides an executable ready to use. Most users would trust and run the executable based on the publisher's reputation. More cautious users may opt to review the source code. However, there is no guarantee that the provided executable comes directly from the published source code. R-B allows testing whether a provided artifact originates from a given source code. A set of independent verifiers can fetch the source code and rebuild the software following the exact same build instructions, as well as using the same dependencies. They publish whether they obtained the same artifacts and the user can now use this information to decide whether or not to trust the distributed executable. This diagram is a simplified version of Fig.~1 from \cite{reproducible-builds-paper}.}
    \label{fig:r-b}
\end{figure}

\paragraph{Reproducible builds (R-B)} The build process of a software is reproducible, sometimes also denoted as deterministic or verifiable, \emph{if, after designating a specific version of its source code and all of its build dependencies, every build produces bit-by-bit identical artifacts, no matter the environment in which the build is performed}~\cite{reproducible-builds-paper}. A set of independent verifiers can be used to test whether a software is reproducible or not, see Fig.~\ref{fig:r-b}.

In this definition, the build dependencies include the details about the build toolchain. The hard requirement of bit-by-bit identical artifacts may look excessive until one realizes that a single bit is what could separate a safe software from an insecure one (see, e.g.,~\cite{CVE-2002-0083}). Note that R-B is not about making software more secure, insecure software will remain insecure.

\paragraph{Bootstrappable builds (B-B)} Most compilers are written in the same programming language they are compiling. For example, modern versions of \texttt{GNU GCC} are written in \texttt{C++}~\cite{gnu-gcc}. The Java Development Kit (JDK) is written in Java, so to build JDK a pre-compiled version of JDK is required. The goal of bootstrappable builds is to end these loops where untrusted compilers have to be used to obtain newer versions of the same compiler by providing clear bootstrapping paths, using compilers written in other languages, and minimizing or eliminating the amount of required binary artifacts to start a build process.

\paragraph{Diverse Double-Compiling} R-B verifiers are vulnerable to certain compiler attacks, such as ``trusting trust'' attacks\footnote{In a ``trusting trust'' attack a malicious compiler produces corrupted executables, including corrupted versions of the same compiler from its source code, making the situation self-perpetuating.}. Diverse Double-Compiling (DDC) is a technique to detect such attacks. It needs three things: The compiler to be tested, its source code, and a second ``trusted'' compiler. The source code of the compiler is then compiled twice: First with the ``trusted'' compiler, and a second time with the resulted compiler of the first compilation. If the final result exactly matches the original compiler, the compiler has not been tampered with~\cite{Wheeler-paper,Wheeler-PhD-thesis}.

\paragraph{Software Bill of Materials (SBOM)} An SBOM is a nested inventory, a list of all the various components used in building a software, including FOSS and proprietary software, third-party libraries and other dependencies. The SBOM is intended to offer transparency and traceability in the software supply chain, and to help securing it~\cite{CISA-SBOM}.

\section{Origin of reproducible builds}

Together with Dennis Ritchie, Ken Thompson won the 1983 Turing Award \emph{``for their development of generic operating systems theory and specifically for the implementation of the UNIX operating system''}. During the award lecture~\cite{thompson1984trustingtrust} he explained how easy it would be to modify compilers to insert malware in critical software, including the compiler itself. He widely publicized the problem of ``trusting trust'' attacks that had only been briefly explored by Paul A. Karger and Roger R. Schell before~\cite{karger1974}. He already pointed out two of the main problems R-B tries to solve: The issue of trusting binary code whose generation is not fully controlled, and the impossibility of securely running untrusted binary artifacts even after a source-level verification. This lecture seeded the interest in reproducible and bootstrappable builds years later.

In the 90s, the company Cygnus provided commercial support for free software. They made the GNU tools they were supporting fully reproducible, including \texttt{gcc}, \texttt{gdb}, \texttt{binutils} and \texttt{make}. Interestingly, this effort was only revealed 25 years later in the Reproducible Builds Project email list~\cite{cygnus-90s}, and remains the first publicly known case of R-B. From the point of view of Cygnus, binary files that did not match bit-by-bit when generated from the same source code were considered a bad software development practice. During the following years of source code changes without aiming for R-B, some unreproducibility issues which had been fixed in the 90s, were reintroduced and then had to be fixed again.

Explicit interest in R-B as a desired goal was expressed in the Debian email list in the early 2000s~\cite{debian-email-discussion-1, debian-email-discussion-2}. However, no specific actions to make Debian packages reproducible were taken. It was even a common believe that such task was not feasible. What finally sparked interest in R-B was money and privacy concerns from Bitcoin and Tor users. Bitcoin Core was made reproducible in 2011 to ensure that the distributed executable was not modified to steal Bitcoin wallets~\cite{bitcoin-gitian}. Inspired by this work and motivated by the fear of targeted attacks against their users, the Tor Browser started to offer reproducible builds~\cite{tor-blog-deterministic} two years later.

Both Bitcoin and Tor projects showed that achieving R-B was feasible. This motivated Debian developer Jérémy Bobbio (lunar) to organize a public discussion about R-B during DebConf13, the annual Debian Project's developer conference. The session identified three of the most common causes of unreproduciblity while building source code (timestamps, build paths and locales), as well as problems with the Debian toolchain~\cite{debconf13-minutes}. It was decided to create a web page to keep track of all efforts to make Debian packages reproducible.

Although the origin of R-B is tightly connected to Debian, it is now a widely discussed topic in other communities, and in the software industry in general~\cite{google-rb, microsfot-rb}. This manifests in a multitude of talks, workshops, and other events focused on R-B (see, e.g.,~\cite{rb-summits,fossy-talk,bornhack-talk}). Recently, R-B gained further attention after the US government issued an Executive Order on \emph{Improving the Nation's Cybersecurity}~\cite{executiveorder2021}, where SBOM and R-B are explicitly mentioned as a solution to supply chain attacks.

Concerning the toolchain used to ensure reproducibility, early successful attempts by Bitcoin and Tor used sanitized and isolated environments, such as minimal well-defined virtual machines, to remove any uncontrolled inputs to the build system. This is a valid approach but leads to slow building times, excessive technical restrictions on the toolchain used by developers, and it cannot fix unreproducibility caused by non-determinism. A completely different approach is to modify the source code and build tools to ensure that only the relevant inputs (source code, dependencies and build toolchain), and nothing else, can produce changes on the final artifacts. For practical reasons, the modern toolchain to produce and test R-B lies in between these two approaches (see, e.g.,~\cite{archlinux-repro}).

\section{Current state}

Many projects have adopted reproducible builds in the last few years. Some, such as Bitcoin Core, Tor Browser, or the WireGuard Android app~\cite{wireguard-android}, even test that their new releases are reproducible before making them publicly available.

Some of the largest GNU/Linux distributions, such as Debian, Fedora, OpenSUSE or Arch Linux, have made a noteworthy effort to provide reproducible packages in their official repositories, although none of them enforce their packages to be reproducible at the moment. The Reproducible Builds Project~\cite{reproducible-builds-website} itself was born with the goal of making Debian 100\% reproducible. As of November 2023, Debian leads the effort among GNU/Linux distributions to become reproducible with 95.2\% of their \texttt{Bookworm/amd64} packages being reproducible~\cite{debian-repro-integration}. 75.8\% of all Arch Linux packages are reproducible~\cite{arch-repro-integration}. In addition, Arch Linux has developed an independent verification system for binary packages, \texttt{rebuilderd}~\cite{arch-rebuilderd-status}, which allows independent verifiers to test the reproducibility of Arch's and even Debian's repositories. In the future, \texttt{rebuilderd} will also support Alpine Linux.

Security and privacy oriented distributions, Qube OS and Tails, have recently joined the effort. Tails installation images, for example, are fully reproducible~\cite{tails}. Other notable OS such as FreeBSD, NetBSD, Alpine Linux or OpenWrt, have plans to become fully reproducible.

The efforts to achieve R-B have led to thousands of patches involving hundreds of FOSS projects. For example, the GNU GCC compiler introduced additional options to support embedding relative paths in the final artifacts instead of absolute paths. Many projects were modified to support the \texttt{SOURCE\_DATE\_EPOCH} specification~\cite{source-data-epoch-spec}, and improve their build automation files (e.g., \texttt{Makefile}) to remove non-deterministic behaviours.

Actively testing for reproducibility has helped improving the overall quality of software by making the build setup more robust under all kind of environments, as well as uncovering issues that would have otherwise remained unnoticed, sometimes even with security implications (see, e.g.,~\cite{gbrowse-bug}). Fixes and improvements on specific FOSS projects, as well as on the build toolchain, benefit all users independently of the operating system they use.

Nowadays, the most common causes of unreproducibility are well understood, and suggested patches or workarounds to fix them are documented (see, e.g.,~\cite{debian-reproducible-notes, debian-repro-common-problems}). Some of the most frequent ones are mentioned in Table~\ref{table:rb-issues}.

\begin{table}
\centering
\begin{tabular}{lcl}
\toprule
Category & Count (\%) & Identified \& documented issues \\
\midrule
Timestamps & 127 (30\%) & C++ macros, Gzip headers, man pages, PDFs generated with \LaTeX, \\
           &            & docs generated by Doxygen, Maven version files, \dots \\
Randomness & 71  (17\%) & random order in tarball files, random hashes in Cython, \\
           &            & random order of static libraries, randomness in fat lto objects, \dots \\
Paths      & 68  (16\%) & build path captured by gcc, build path captured by Rust, \\
           &            & absolute paths in CMake files generated by Meson, \dots \\
\bottomrule
\end{tabular}
\caption{Summary table containing some of the well-known causes of unreproducibility, from a total of 422 documented and categorized issues in~\cite{debian-reproducible-notes}, as of November 2023. For actual source code examples also check ``The Unreproducible Package'' repository~\cite{unreproducible-pakcage}. This classification also applies to other GNU/Linux distributions, although some of the technical solutions or recommended practices may differ.}
\label{table:rb-issues}
\end{table}

Tools are available to assist with the root cause analysis of unreproducible artifacts. The most widely used is \texttt{diffoscope}~\cite{diffoscope}. It transforms multiple binary formats into more human-readable forms and presents the results as a \texttt{diff} text or html, allowing a better comparison. Embedded times and paths are highlighted, while it can also help debugging more complex non-deterministic issues such as embedded randomness, uninitialized memory or address space layout randomization. Despite some recent progress in automating these tasks~\cite{zhilei-automated-patching}, achieving R-B remains a manual and highly time-consuming process.

\section{Challenges}\label{sec:challenges}

\paragraph{Awareness} Despite the efforts of both the Reproducible Builds Project~\cite{reproducible-builds-website} and the Bootstrappable Build Project~\cite{bootstrappable-website}, these topics still remain unknown to many developers. Raising awareness about software supply chain attacks and the benefits of R-B and B-B will improve the situation and further accelerate the goal of being able to run a fully reproducible operating system. Aware programmers are less likely to introduce new sources of non-deterministic behaviour in their programs that could cause regressions in the reproducibility status.

\paragraph{Policy changes} Even though most GNU/Linux distributions working on R-B have achieved success ratios above 90\% in their official repositories, some of the remaining unreproducible packages will be challenging to fix since they involve multiple changes upstream. Unmaintained unreproducible software will never be fixed, and a small fraction of the remaining software will not be reproducible by decision of their developers~\cite{ghostscript-r-b}. Achieving a 100\% reproducible ratio for GNU/Linux distributions with large repositories, such as Debian, will only be possible together with a clear decision of not packaging unreproducible software. Such decision will be controversial, and may lead to new forks like it happened in the past~\cite{devuan}.

\paragraph{Archive infrastructure} Continuously testing for package reproducibility involves downloading old versions of dependencies that may no longer exist in the stable repositories. Debian and Arch Linux maintain their own archives servers~\cite{debian-snapshot, archlinux-archive} that allow such testing. Debian Snapshot, for example, includes all packages since 2005. Arch Linux Archive only maintains packages for a few years\footnote{As of November 2023, the oldest Linux kernel available in the Arch Linux Archive is version 4.20.10 packaged in February 2019.}, but older packages are moved to the Internet Archive~\cite{internet-archive} before being deleted. Maintaining this kind of infrastructure requires time and effort, and comes with a huge cost that smaller projects cannot afford.

\paragraph{Independent rebuilders} R-B are only useful with independent verifiers continously rebuilding the software, allowing users to query whether a distributed software is reproducible or not (see Fig.~\ref{fig:r-b}). This is lacking at the moment as only a few testing platforms exist, most of them maintained by the Reproducible Builds project. Ideally, in the future, easy to setup rebuilders will allow users and companies to run their own reproducible tests and report their results. In addition, protocols have to be developed to allow users to quickly decide if a binary is to be trusted.

\section{Our contribution to the reproducible builds effort}

\subsection{Hochschule Luzern verifier for Arch Linux reproducibility}

One of the key open issues to fully benefit from R-B is the lack of independent verifiers that continuously test packages, or other artifacts, to check if they are reproducible (see Sec.~\ref{sec:challenges}). Only with enough verifiers confirming reproducibility, we can trust that the distributed artifacts originate from a given source code.

We have set up a server running \texttt{rebuilderd}~\cite{rebuilderd} to test the reproducibility of Arch Linux packages in the official repositories. The reason we have chosen to use this distribution is twofold: Arch Linux provides packages with minimal distribution-specific changes with respect to upstream versions, and the package build description files are \texttt{bash} scripts, easy to read and understand even for people without prior packaging experience.

\begin{figure}[H]
\centering
\begin{tikzpicture}	[line width=1pt]
    \path[clip] (-1.45,-0.1) rectangle (11.3,2.35);
    \draw [draw=black](0.6,0) rectangle (7.5,2.3);
    \node(archmirror)[rectangle, align=center, anchor=center, draw=black, minimum height=1.0cm, minimum width=1.0cm, inner sep=2pt] at (-0.5,1.15) {Arch Linux\\mirror};
    \node(syncdeamon)[rectangle, align=center, anchor=center, draw=black, minimum height=1.7cm, minimum width=1.2cm, inner sep=2pt] at (1.5,1.15) {sync\\daemon};
    \draw[->,color=black] (syncdeamon) -- (archmirror);
    \node(rebuilderd)[rectangle, align=center, anchor=center, draw=black, minimum height=0.5cm, minimum width=3cm, inner sep=2pt] at (4,0.5) {rebuilderd};
    \node(w1)[rectangle, align=center, anchor=center, draw=black, minimum height=0.8cm, minimum width=1cm, inner sep=2pt] at (3, 1.6) {worker\\\#0};
    \node(w1)[rectangle, align=center, anchor=center, minimum height=0.5cm, minimum width=1cm, inner sep=2pt] at (4,1.6) {\ldots};
    \node(wn)[rectangle, align=center, anchor=center, draw=black, minimum height=0.8cm, minimum width=1cm, inner sep=2pt] at (5,1.6) {worker\\\#n};
    \draw[<->,color=black] (3,0.75) -- (3,1.15);
    \draw[<->,color=black] (5,0.75) -- (5,1.15);
    \draw[<->,color=black] (2.1,0.5) -- (2.5,0.5);
    \node(webserver)[rectangle, align=center, anchor=center, draw=black, minimum height=1.7cm, minimum width=1.2cm, inner sep=2pt] at (6.5,1.15) {NGINX\\web\\server};
    \draw[<->,color=black] (5.5,0.5) -- (5.9,0.5);
    %\node(deb)[align=center, anchor=center] at (1.0,2.1) {\includegraphics[width=0.5cm]{images/openlogo-100.png}};
    \node(screenshot)[rectangle, align=center, anchor=center, draw=black, minimum height=1.75cm, minimum width=2.5cm, inner sep=2pt] at (9.45,1.1) {\includegraphics[height=1.75cm]{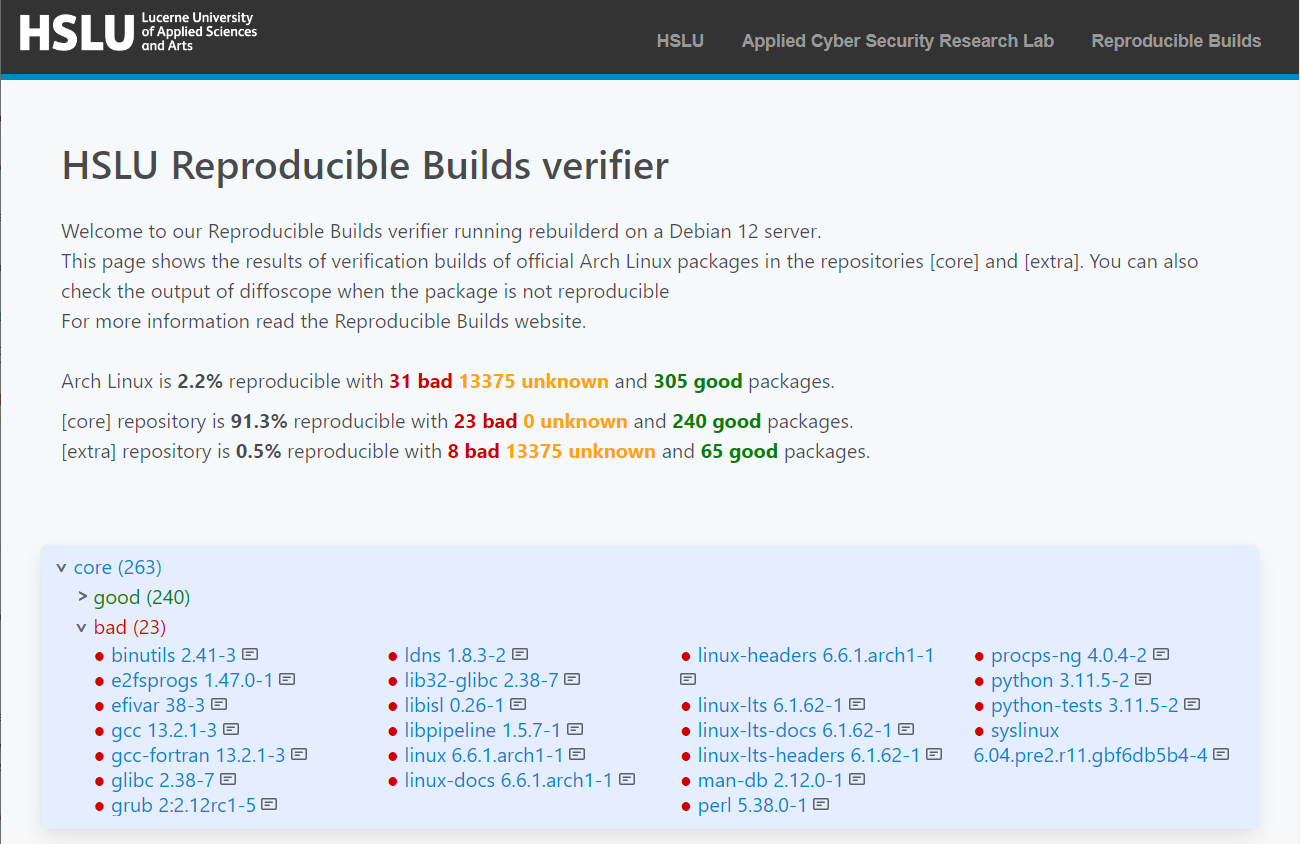}};
    \draw[<->,color=black] (webserver) -- (8,1.15);
    \end{tikzpicture}
    \caption{The \texttt{rebuilderd} daemon requires few resources and can be run on a small cloud instance (e.g., 1 vCPU and 1 GB of RAM). Our independent verifier (sync daemon and a small worker) runs on a modest refurbish PC with an Intel i7-4790K CPU and 16 GB of RAM. Workers need at least 32 GB of RAM to be able to build all the official packages. A significant speedup on large builds (e.g. linux kernel, web browsers, etc.) can be achieved with at least 64 GB of RAM and mounting the worker's directory as \texttt{tmpfs}. For the initial setup, two VMs with 16 vCPU, 64 GB of RAM and 200 GB of SSD storage were used as workers.}
    \label{fig:rebuilderd}
\end{figure}
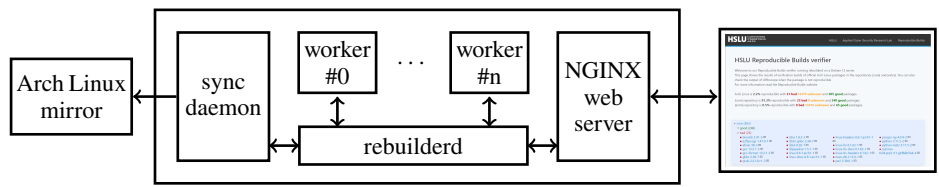

The verifier works in the following way: A sync daemon checks and downloads the latest packages from the official Arch Linux repositories every 5 minutes. The \texttt{rebuilderd} daemon splits the work among the active workers, that can be dynamically deployed based on the workload, and exposes an API to query the status of the packages: GOOD are reproducible, BAD failed to build or are not reproducible, and UNKWN have not been checked yet. A website available at \url{https://reproducible.crypto-lab.ch} shows the status of the packages, build logs, \texttt{diffoscope} logs for unreproducible builds and attestation logs for reproducible packages. A simplified diagram of our setup, as well as the specs of the hardware used can be seen in Fig.~\ref{fig:rebuilderd}. A list of all independent \texttt{rebuilderd} instances can be found in the Arch Linux wiki~\cite{rebuilderd-wiki}.

Packages marked as reproducible are not tested again until a new version becomes available in the official repositories. Unreproducible packages, however, are requeued endlessly by \texttt{rebuilderd} after some delay. This delay is determined by the number of attempts times a configurable base amount. We increased the value of the base delay from 24 hours to 1 week in order to avoid delaying the test of new versions of packages over unreproducible ones.

\subsection{Fixing unreproducible Arch Linux packages}

While exploring some of the unreproducible packages detected by our instance, we uncovered and reported a packaging bug in how all the \texttt{certbot} packages are created in Arch Linux~\cite{archlinux-bug-80266}. Certbot~\cite{certbot} is the official tool to obtain TLS certificates from Let's Encrypt~\cite{letsencrypt}. It can be considered critical software since it is currently used by millions of users to manage certificates and malicious versions could potentially leak the private keys. All certbot-related packages are considered reproducible in Debian's testing infrastructure~\cite{debian-reproducible-certbot}, so there are no technical reasons why they should not be reproducible in Arch Linux as well. The underlying problem affecting the 16 packages was the workflow followed to update the package to a new version.

Arch Linux uses a \texttt{bash}-alike script called \texttt{PKGBUILD}~\cite{PKGBUILD} to describe the instructions to create a package from source code, and stores an SBOM in a key-value formattted file called \texttt{.BUILDINFO}~\cite{BUILDINFO} inside the generated package.  Dependencies required to run, build and test a package are specified in the \texttt{PKGBUILD} variables \texttt{depends}, \texttt{makedepends}, and \texttt{checkdepends}, respectively. They are all fetched and installed in the building machine before any function is executed. A minimal \texttt{PKGBUILD} must contain at least the \texttt{package()} function, which is used to install the relevant files into the packaging directory. Optionally, the functions \texttt{prepare()}, \texttt{pkgver()}, \texttt{build()} and \texttt{check()} can also be used. Relevant to understand the issue we uncovered is the \texttt{pkgver()} function, which is used to update the \texttt{pkgver} variable where the version of the package is stored. In this particular case, a fixed and manually defined git commit is used by this function, which determines the associated tag version and uses it to update \texttt{pkgver}. The issue was that \texttt{pkgver} had been previously used to determine the version of one of the dependencies, \texttt{python-acme}. This is accurately recorded in the SBOM of the package, causing a mismatch between the required version by the \texttt{PKGBUILD} script and the actual version used to produce the official package.

To avoid similar problems in the future, Arch Linux build toolchain could be patched to abort in situations where a function can modify the building description (\texttt{PKGBUILD}) after the dependencies have been fetched and installed.

\subsection{Fixing unreproducible source code}

Fixing unreproducible packages in a specific distribution is a useful contribution to the R-B efforts. Even better is fixing the causes of unreproducibility directly in the source code of the upstream projects, since this has the potential to fix unreproducible packages for all distributions working towards that goal. Using the insights from our rebuilder instance, we have further discovered an issue in the building scripts of \texttt{fwupd}~\cite{fwupd}. Our motivation to make \texttt{fwupd} reproducible is that it is closely related to securing software security chains, as it is the standard software to update firmware in Linux devices.

In this case, the very well-known issue of an embedded timestamp in the header of a compressed file (see Table~\ref{table:rb-issues}) was causing the unreproducibility. The reason why this was not caught earlier is probably because the file was generated using a Python script and \texttt{gzip} from the Python Standard Library~\cite{python-gzip} instead of the standard GNU tool \texttt{gzip}~\cite{gnu-gzip}. We submitted an upstream patch to fix the issue, which was quickly accepted~\cite{fwupd-fix}. This will make the package in Arch Linux reproducible~\cite{archlinux-bug-80271}, as well as in any other distribution packaging the latest release of \texttt{fwupd} once it becomes available.

\section{Conclusions}

In recent years, a clear trend has emerged aimed at enhancing the security of software supply chains. Some countries are actively advocating, and in some cases mandating, the implementation of software bill of materials (SBOM). Reproducible builds (R-B) is the only solution to validate these SBOMs through independent verification. Bootstrappable builds (B-B) serve as a definitive protective measure against ``trusting trust'' attacks during the build process. The combination of both B-B and R-B ensures a clear and unequivocal linkage between source code and binary code. When complemented with routine source code audits, these measures effectively mitigate the risk of supply chain attacks.

Despite the great number of achievements of R-B during the last ten years, there are still many open challenges to achieve the goal of running 100\% reproducible operating systems. First and foremost, an independent network of rebuilders and verifiers is needed. Our instance of an independent verifier supports this cause. In addition, all the software running these verifiers should be itself reproducible, which is still an open issue. Second, many specific issues in the build process and upstream source code still need to be fixed to make all packages reproducible. We have reported on both types of issues, uncovering a packaging issue affecting the Arch Linux build process, and fixing directly the source code of a crucial software to make it reproducible.

We finally hope with this paper to encourage others to do the same and join the reproducible builds effort.

\section{Acknowledgments}
We thank the core team of the Reproducible Builds Project, and the Arch Linux developers and package maintainers active in the \#archlinux-reproducible IRC channel for valuable insights, especially kpcyrd, the author of \texttt{rebuilderd}, and Jelle van der Waa, the maintainer of the official Arch Linux reproducible status website.

This work was supported by SNSF Practice-to-Science grant no.~199084.

\newpage
\printbibliography
\end{document}